\documentclass[preprint,12pt]{aastex}
\usepackage{epsfig}
\usepackage[dvips]{color}
\def\rd{Di\thinspace Stefano}

\shorttitle{Close planets with microlensing}
\shortauthors{Di Stefano} 

\begin{document}


\title{Discovering habitable Earths, hot Jupiters and other close planets with microlensing}

\author{R. Di\thinspace~Stefano}
\affil{Harvard-Smithsonian Center for Astrophysics, 60
Garden Street, Cambridge, MA 02138}

%
%

\begin{abstract}
Searches for planets via gravitational lensing have focused on cases in which
the projected separation, $a$,  between planet and star is comparable to the 
Einstein radius, $R_E.$ 
This paper 
considers smaller orbital separations and demonstrates that 
evidence of close-orbit
planets can be found in the    
low-magnification portion of the 
 light curves generated by the central star. 
We develop a protocol to discover hot Jupiters as well as 
Neptune-mass and Earth-mass planets in the
stellar habitable zone.
When planets are not discovered, our method can be used to quantify the
probability that the lens star does not have planets within specified ranges of
the orbital separation and mass ratio. 
Nearby close-orbit planets discovered by
lensing can be subject to follow-up observations to study the newly-discovered
planets or  to
discover other planets orbiting the same star. 
Careful study of the low-magnification portions of lensing light curves
should produce, in addition to the 
discoveries of close-orbit planets,
definite detections of wide-orbit planets through the discovery
of ``repeating'' lensing events. We show that  
events exhibiting extremely high magnification can  
effectively be probed for planets in close, intermediate, and wide distance 
regimes simply by adding several-time-per-night monitoring in the
low-magnification wings,    
possibly leading   
to gravitational lensing discoveries of 
multiple planets occupying a broad range of orbits, from close to wide,  
in a single planetary system.
\end{abstract}

\maketitle

\def\rd{Di\thinspace Stefano} 

\section{Introduction}

Microlensing planet searches have been directed toward discovering 
planets in orbits whose size is comparable to the size of the 
Einstein radius, $R_E$, 
of the central star. Here we study the detectability 
of planets in much closer orbits. This is necessary, 
because we now know that many planets are in 
such close orbits\footnote{http://exoplanet.eu/catalog.php}. 
We demonstrate that ground based surveys for lensing events
can detect a wide range of close-orbit planets,
including ``hot Jupiters'' orbiting 
sun-like stars, and even
Earth-mass planets in 
the habitable zones of M dwarfs. We also discuss the role of space missions. 

We were led to the study of close-orbit planets by
 work to determine whether or not planets in the 
habitable zones of nearby stars could produce 
detectable lensing signatures 
(\rd\ \& Night 2008). 
The standard planetary-lensing scenario is most sensitive to planets
with $\alpha=a/R_E$ in the range $0.5-2,$
 where $a$ is the
orbital separation (Mao \& Paczy\'nski 1991; Gould \& Loeb 1992).
In many cases, this corresponds to the region 
beyond the snowline (e.g., Sumi et al.\, 2010), while 
  the habitable zone 
is closer to the star.
We found that lensing observations {\sl would}
 be able to detect planets
orbiting within the habitable zones of nearby M dwarfs (\rd\ \& Night 2008).
Here, we demonstrate the importance of lensing by close-orbit planets 
for a broader range of systems: the planets may be in the habitable
zone, closer in, or farther out. The planetary systems may be
within a few tens of pc or much farther away. The host stars may
be M dwarfs, brown dwarfs, or much more massive stars.    

In \S 2 we study the incidence of close planets. In \S 3 we study the
expected signatures. We sketch the steps needed to mount successful
programs to discover these planets in \S 4.
This paper provides a guide to the 
discovery and analysis of lensing
light curves generated by close-orbit planets.

\section{Evidence for Close-Orbit Planets}

As this text is being written, 696 planets are listed in the {\it Interactive
Extra-solar Planets Catalog}\footnote{http://exoplanet.eu/catalog.php}.
Most of these planets ($\sim 645$)
have been detected by radial velocity (RV) studies of the host star.
The second-largest group ($\sim 185$) have been detected through planetary
transits. There is some overlap between these two groups. Twenty six
planets have been detected by imaging, and 12
have been discovered through planet-lens signatures detected 
during gravitational 
lensing events in which the host star serves as the primary lens. 

Searches for planet-lens events have focused on the case in which 
the orbital separation at the time of the event, $a$, is comparable in size to
the Einstein radius, $R_E,$ of the central star: 
$0.5 < \alpha < 2,$ with $\alpha=a/R_E.$ To determine the effect 
this has on
planet discovery, 
we treat each exoplanet central star as a potential lens, and
determine the value of $\alpha$ for each of the discovered planets.
To do this calculation, we need to 
compute the value of $R_E$ for each central star. 
\begin{equation}  
R_E= 1.01\, {\rm AU}\, \Bigg[\Big(\frac{M_\ast}{1\, M_\odot}\Big)\, 
                         \Big(\frac{D_L}{125\, {\rm pc}}\Big)\, 
	                 \Big(1-\frac{D_L}{D_S}\Big)\Bigg]^\frac{1}{2}, 
\end{equation} 
 
To compute $R_E,$ we need the star's mass, 
$M_\ast,$ and its
distance $D_L$ from us. The value of $R_E$ also depends on
the distance, $D_S,$ to the source that would be lensed.
Because most of the central stars do not have a 
bright source located directly behind them, the value of $D_S$ is
not determined.   
To compute $R_E$ we must therefore  
make some assumptions about the
ratio $D_L/D_S$.
Here we first consider
lenses for which 
$D_L/D_S << 1$; this allows us to ignore the last factor in Equation 1.
The physical meaning of this assumption is that we are focusing on
nearby planetary systems or on planetary systems with source stars located
much farther from us.  
We will mention a second case in \S 4, with $(D_S-D_L)/D_S << 1,$ generally
corresponding
to lens stars very close to the lensed source. 

Making these assumptions, we can compute
$\alpha=a/R_E,$ for each known exoplanet planet
whose semimajor axes $a$ has been measured. 
Figure 1 shows the results for all planets for which
we have estimates of $a, M_\ast, M_{pl}, D_L,$ and the  orbital period,
$P.$ We have defined $q=M_{pl}/M_\ast.$ 
The two dashed lines in each panel enclose the region 
$0.5  < \alpha < 2$.    
This is the region for which microlensing
planet searches have been primarily directed.
Only about 1/3 of the known planets fall in this region.
Even among systems with $\alpha$ in this range,
planets will be discovered for only a small fraction. This is because the
path of the source behind the lens must be favorable for planet
discovery and also because the sensitivity and sampling cadence 
must be well suited to planet discovery. 

It is therefore advantageous to extend lensing planet searches to both 
smaller and larger values of $\alpha.$ Larger values have been considered
in some detail (Di\thinspace Stefano \& Scalzo 1999a, 1999b; Han 2009). 
We will discuss them briefly in \S 4. The major part
of this paper is devoted to studying planet detection of smaller values of 
$\alpha.$  

From Figure 1, we see that, judging by the
planetary systems already known,
 smaller values of $\alpha$ are expected for 
a wide range of stellar masses, planet masses, and distances $D_L$.
One variable that displays a trend with $\alpha$ is the
orbital period, $P$.
\begin{equation} 
P=71.3\, {\rm days}\, \Bigg(\frac{\alpha}{\frac{1}{3}}\Bigg)^{\frac{3}{2}} 
         \Bigg(\frac{M_\ast}{M_\odot}\Bigg)^{\frac{1}{4}} 
\Bigg(\frac{D_L}{125\, {\rm pc}}\, \Big(1-\frac{D_L}{D_S}\Big)\Bigg)^\frac{3}{4}  
\end{equation} 
Thus, smaller values of $\alpha$ are
associated with shorter orbital periods. 
In the next section we 
will study the geometry of the isomagnification
contours of close-orbit planetary systems, and will find 
that there is a small region exhibiting deviations from the
point-lens form at large distances ($u>1\, R_E$) from the
center of mass. This region rotates around the center of mass
at the orbital period. The region of deviation can therefore
rotate into the path of the source track, increasing the probability
of detection.

If $v$ is the relative transverse motion, the proper motion is
\begin{equation}  
\mu = 0.0338^{\prime\prime}{\rm yr}^{-1} 
\Bigg(\frac{v}{20\,  {\rm \frac{km}{s}}}\Bigg)\, 
\Bigg(\frac{125\, {\rm pc}}{D_L}\Bigg) 
\end{equation}  
The Einstein angle is
\begin{equation}  
\theta_E = 0.0081^{\prime\prime} 
 \Bigg[\Big(\frac{M_\ast}{1\, M_\odot}\Big)\, 
                         \Big(\frac{125\, {\rm pc}}{D_L}\Big)\, 
	                 \Big(1-\frac{D_L}{D_S}\Big)\Bigg]^\frac{1}{2}, 
\end{equation}  
Define $\tau_{E,1}$ to be the  time taken for the source-lens separation
to change by an angle equal to the Einstein angle. 
For nearby lenses, 
$\tau_{E,1}\approx \theta_E/\mu$, and its value can be comparable to
the value of $P,$ when $\alpha$ is small.    
For, example, a solar mass lens
at $125$~pc will have $\tau_{E,1}\approx 88$~days if
$v=20$~km~s$^{-1}.$ If the detection
limit is $2\%$, the event may be detectable during the time taken to
cross through $6\, R_E.$ For a range of lens masses and distances, 
several orbits may occur during a lensing event.   

\section{Close-Planet Magnification Geometry and Light Curves}

\subsection{The Role of $\alpha$} 

When the projected distance between planet and star is 
significantly smaller
than $R_E,$ then at distances larger than $R_E,$ the system 
is well approximated by a point-lens of total mass
equal to the sum of the stellar and planet masses.  
Nevertheless, there are small deviations from the point-lens form.
To study these deviations we consider lenses with planets, each planet
characterized by the mass ratio $q=M_{pl}/M_\ast,$ and the separation $\alpha.$ 

We begin by considering the magnification geometry for a fixed value of $q$ by
computing 
the magnification around each of
 a sequence of
concentric rings, centered on the center of mass. 
If the system were a point mass, we would expect that 
each ring of radius $r$ would have a single magnification, $A(r)$, whose value would be
equal to $(r^2+2)/(r\, \sqrt{r^2+4})$. 
For each ring, we computed the difference, $\Delta,$ between the 
maximum and minimum value of the magnification.  
 Deviations from symmetry are
associated with values of $\Delta$ that differ from zero.
The top panel of 
Figure 2 shows $\Delta$ as a function
of $r$ for a lens with $q=0.001$.  
The most striking feature of this panel is the sequence of $9$ peaks.
Each corresponds to a single value of $\alpha,$ and the peak occurs at 
$r = R_{\alpha} = \frac{1}{\alpha} -\alpha.$ 
The black peak on the 
right corresponds to $\alpha=0.1.$ The value of
$\alpha$ increases by $0.05$ 
for each peak toward the left, to $\alpha=0.5$
for the left-most peak. 
For values of $\alpha$ near or above $0.5$, lensing by planetary systems
has been well studied; the black curve ($\alpha=0.5$) shows that there are
deviations from the point-lens form larger than $1\%$ over a wide range of
values of $r$. For smaller $\alpha,$ however, the deviations in the region
$r>1$ are small, except in the peaks, where they can be as large as several 
tens of percent.  
We would derive a similar pattern for other
values of $q$. In fact, the locations of the peaks would be identical. The width
of the peaks would be larger (smaller) for larger (smaller) values of $q$. 

To see why non-linear effects become evident 
at the star-planet separation $r=R_\alpha=\frac{1}{\alpha}-\alpha>1,$  
consider the image geometry
for the simplified case of a point source, located a 
distance $R_{\alpha}$ from a point lens.
If the $x$ axis connects the lens and  source, then at a value of $x$ equal to
$- \alpha,$ there will be a negative parity image of the source.
When a planet happens to lie near this point,  
its influence on the total magnification will be enhanced. 

Now consider a planetary system with the center of mass at the origin and the 
planet at $x=- \alpha.$  
There are two tiny caustics located along the circle of radius
$R_\alpha;$ one appears at a positive value of $y$ and one
appears at a negative
values of $y.$ The caustics themselves are too small to play a significant role,
but they serve as a convenient way to locate the regions in the lens plane 
within which the magnification deviates from the point-lens form. 
When the source happens to lie behind one of these ``perturbed'' regions,
the light curve will exhibit features that signal the presence of the planet.

\subsection{The role of the mass ratio, $q$} 

As shown in \S 3.1, the value of $\alpha$ determines the distance from
the center of mass of the
region with isomagnification perturbations. In this subsection we 
show that the size of these perturbed
regions is determined by the mass ratio, $q$.
The alterations in the isomagnification contours are shown in the
bottom-right panels of Figures 3, 4, and 5.  With
$\alpha=1/3$ in all three cases, these panels differ from
each other only in the value of $q$, which is $1.25 \times 10^{-3}, 
2.0\times 10^{-4},$ and $1.2\times 10^{-5}$ in Figures~3, 4, and 5,
respectively. 
These figures demonstrate that the size of the perturbed regions
is smaller for smaller values of $q$. 

The figures for the two smallest values of $q$ correspond to a Neptune-mass 
planet and an Earth-mass planet, respectively, orbiting a star with
$0.25\, M_\odot.$ 
In each case, the lower-right panel zooms in on the perturbed
region, to reveal that, in a small region of the annulus around $R_\alpha$,
isomagnification contours from larger values of $r$
are pulled in to smaller values of $r$, with the contours from
smaller $r$ pushed out on either side.  
When a source with larger transverse than radial speed
 passes behind this region, the magnification will deviate
upward from the point lens form, then downward and up again before
descending back to the point-lens value. As shown in the top panel
of all three figures and in the portion of the light curve
shown in the lower left-hand panel of each figure, this  
characteristic ``up-down-up-down'' form of the light curve   
is exhibited when both $\alpha$ and $q$ are small. 
 
The isomagnification contours in the 
lower-right-hand panels for the Neptune-mass and Earth-mass planets
exhibit small closed curves, which  
enclose
caustics. The caustics are tiny and their positions are not marked here;  
in fact, the caustics do not play an important role in the light curve
deviations. The light curve deviations
are dominated instead by the more subtle affects
associated with the perturbations of the low-magnification
isomagnification contours. Nevertheless, the positions
of the caustics, which can be vanishingly small, provide a convenient
way to measure the size of the perturbed region.

We define $\Delta y_c$ to be the straight-line 
distance between the tiny caustics discussed above, 
expressed in units of $R_E.$ 
We compute a normalized separation, $\Delta\, Y_{norm},$  
by dividing $\Delta y_c$ by $C_\alpha=2\, \pi\, R_\alpha$, 
the 
circumference of a circle of radius 
$R_\alpha.$ 
Consider the bottom panel of Figure 2. 
The variable along the vertical axis is
the logarithm of the normalized separation, $\Delta\, Y_{norm}$; 
it is plotted against $log_{10}(q).$   
There are $5$ colored curves for values of $\alpha$ ranging from $0.10$ to
$0.33$; these curves are almost indistinguishable. Moving to wider orbits,
the green 
curve for $\alpha=0.40$   
can be distinguished from the others,
but it is close to  them. All in all, there is
very little alpha dependence, indicating that the linear dimensions
of the perturbed area depend primarily on the value of $q$. 
The curves for small $\alpha$ and small $q$
 are well approximated by the equation: 
$log_{10}(\Delta\, Y_{norm}) = 0.5\, log_{10}(q) - 0.2.$
Thus, the physical separation, expressed in units of $R_E$,
can be expressed as a product of a factor that depends only
on $\alpha$ and one that depends only on $q$: 
$\Delta\, y_c = 2\, \pi\, R_\alpha \Delta\, Y_{norm}(q).$  

Figures 4 and 5 clearly show that the perturbed region
is larger than the distance between the centers of the closed curves,
which is an approximate  measure of the separation  
$\Delta\, y_c$ between caustics. Let $L(\alpha, q)$ represent the 
linear dimensions of the perturbed 
region, expressed in units of $R_E$. 
On an empirical level, the size of the region is determined by the
size of the smallest deviations that can be reliably detected
for any given observational scheme. If deviations like those shown
in the light cures in the top panels of Figures 3 through 5\footnote{
We note that, in order to generate these particular light curves,
we used face-on circular orbits. The general theory applies to orbits of
all orientations and eccentricity; the value of $\alpha$ and the geometry of the
isomagnification contours are then time dependent. Nevertheless, if the 
time duration of the deviations from the point-lens form is primarily
determined by the value of the orbital period, the basic shape of individual
deviations will have the same characteristics as shown here.} 
 are detectable, then  
we find, empirically, that 
$L(\alpha, q) \approx 2.5\, \Delta\, y_c.$ 

For the three cases shown in Figures 3, 4, and 5, the linear 
dimensions, $L(\alpha, q),$ are 
approximately $0.93\, R_E, 0.37\, R_E,$ and $0.09\, R_E,$ 
respectively.
In the absence of orbital rotation, the event rate would be 
proportional to these linear dimensions. The event durations would 
be equal to the time taken for the relative lens and source positions
to change by $L(\alpha, q).$ Let this time be denoted by $T_{transverse}.$
\begin{equation}     
T_{transverse}=21.6\, {\rm days} \, 
\Big(\frac{L(\alpha, q)}{0.25}\Big)
\Big(\frac{R_E}{1\, {\rm AU}}\Big)
\Big(\frac{20\, {\rm km/s}}{v}\Big)
\end{equation}     
In most cases, however, the events will be significantly shorter,
because orbital motion plays an important role. In addition, orbital
motion increases the likelihood that a detectable event will occur.
In the case in which $T_{transverse} > P,$ the probability that a detectable
event will occur is unity. 

\subsection{Event Probabilities and the Role of Orbital Motion} 

Particularly in cases with
$\alpha < 0.5,$ the orbital period can be comparable to or even shorter
than  
$T_{traverse}$ the time taken for the source  
to traverse a distance $L(\alpha, q).$ 
In such cases, the
orbital motion is very likely to rotate the perturbed region in front
of the source. The probability is  $T_{traverse}(q)/P_{orb}$ when
 $T_{traverse}(q)< P_{orb}$ and is unity otherwise. When the probability 
is larger than unity, deviations repeat on a time scale roughly
equal to $P_{orb}$. If we are monitoring the system
frequently enough to catch a deviation in progress, our chance of
seeing the deviation can be $100\%$ if the planet exists. 

The time duration of a deviation from the point-lens form is
\begin{equation} 
T_{dev} = \frac{L(\alpha, q)}{2\, \pi\, R_\alpha} P_{orb}=
2.5\, P_{orb}\,  10^{(0.5\, Log_{10}(q) - 0.2)}   
\end{equation} 
In the cases shown in Figures 3, 4, and 5, this produces deviations of
durations $1.1$~days, $0.56$~days, and $0.14$~days, which 
compare well with values of $T_{dev}$ shown in
the bottom-left panels of each figure.

\subsection{Hot Jupiters}

The first exoplanet to be discovered orbiting a sun-like star was 
51~Peg~b (Mayor \& Queloz 1995),
a planet with $m\, sin(i) \approx 0.5\, M_J$ in a 4-day orbit. At present,
there are 155 known exoplanets with semimajor axis smaller than $0.5$~AU and
with $m\, sin(i)$ between $0.5\, M_J$ and $10\, M_J.$ These planets are generally
referred to as ``hot Jupiters''. The value of $m\, sin(i)$ is larger than
($1\, M_J, 2\, M_J, 3\, M_J$) in ($96, 57, 31$) cases, respectively.    

The planet corresponding to Figure~3 is a hot Jupiter. Its mass is
equal to that of Jupiter, and it orbits a star of $0.8\, M_\odot;$ 
$\alpha=1/3$. With $D_L=25$~pc; $D_S=8$~kpc, we find  
$R_E=0.4$~AU. Thus, the semimajor axis is $a= \alpha\, R_E = 0.13$~AU, and
the orbital period is $20$~days. With a transverse speed of $20$~km/s,
the time taken to cross $R_E$ is $\tau_{E,1}=35$~days. The duration of the
deviations is just over a day, and the deviations repeat. This example
demonstrates that hot Jupiters can be found through their influence on the
low-magnification portion of lensing light curves.  Note that if, for the
same lens star and orbit, the planet had a mass of 
($3\, M_J, 6\, M_J, 10\, M_J$), then $T_{dev}$ would be
($1.9$~days, $2.7$~days, $3.5$~days).         

Lensing provides a potentially important complement to the radial-velocity
and transit studies that have already been discovering hot Jupiters.
It allows planet discovery even if the central star is too dim for detailed
spectral studies, and for {\sl all} orbital inclinations, in contrast
to transit studies. Furthermore, lensing provides a direct measure of the 
lens mass, at least in cases in which the mass of the central star can be 
determined. Furthermore, as we show below,
lensing searches for hot Jupiters can be effective
for nearby stars, allowing detailed follow-up studies, and also for distant
stars.  

An important issue is whether the discovery of hot Jupiters, or placing
reliable limits on their presence around lens stars, can be accomplished on a
regular basis. To answer this, we consider a condition sometimes used to define
the boundary between hot Jupiters and planets farther out: $a < 0.5$~AU.  
\begin{equation} 
 \alpha \, R_E < 0.5\, {\rm AU} 
\end{equation}
Consider background sources located in the Galactic Bulge ($D_S=8$~kpc). 
If the central star is a low-mass dwarf, with $M_\ast\sim 0.1\, M_\odot,$
then the condition above holds for all values of $D_L.$  
For stars of $0.25\, M_\odot, 0.5\, M_\odot, 0.75\, M_\odot, 1.0\, M_\odot,$
the condition holds for $D_L< 1.4$~kpc, $600$~pc, $400$~pc, and $300$~pc,
respectively, and also for $D_L> 6.6$~kpc, $7.4$~kpc, $7.6$~kpc, and $7.7$~kpc,
respectively.
Since a large fraction of the lenses are located in the Bulge itself or
else within a kpc of Earth (Di\thinspace Stefano 2008a, 2008b), searches 
for hot Jupiters in the low-magnification portion of lensing light curves
are feasible. 
 
\subsection{Neptunes and Earths in the Habitable Zone}

In both Figures 4 and 5, the central mass is $0.25\, M_\odot,$ and $D_L$ is
$50$~pc. This yields, $R_E \sim 0.32$~AU. As in Figure 3, $\alpha=1/3.$ The
separation between planet and star is $\sim 0.1$~AU. The flux incident on
the planets associated with both figures is, therefore, about $78\%$ the
flux received by Earth from the Sun, and these planets are in or near the
habitable zone. The orbital periods are about $25$ days, and
the transverse speed was $15$~km~s$^{-1}$.

Figures 4 and 5 show the light curve, the characteristic form of the
deviations from the point-lens case, and the perturbations of the 
isomagnification contours for the Neptune-mass and Earth-mass planets,
respectively. 
The shorter duration of the deviations 
for the lower mass planets means that higher-cadence
sampling would be required to fully resolve them.
Nevertheless, the general up-down-up-down form 
is clear in all three cases. 
Furthermore, 
the light curves shown in the top panels of both figures
indicate that the orbital period may be recoverable in cases such as these.

If planets are not uncommon in the habitable zones of their stars, then
studying the low-magnification portions of lensing light curves for
evidence of planets in $\alpha < 0.5$ orbits is an effective way to
discover them, both for nearby and distant planetary systems. 
Di\thinspace Stefano \& Night (2008) computed the range of 
stellar-lens masses and distances for which a planet with a given value
of $\alpha$ would be in the habitable zone. Their results indicate that
close-orbit planets in the habitable zone could be detected for a wide range of
values of $D_L.$ For example, $\alpha$ is smaller than $0.5$ for planets in the
habitable zone of a $\sim 1.5\, M_\odot$ star, with $800$~pc $< D_L < 7200$~pc.

\section{Successful Observing and Analysis Strategies} 

Lensing associated with close orbit planets is a new frontier.
Fortunately, the ongoing monitoring programs can allow us to 
begin exploring
this frontier in the immediate future. 
Below we summarize the relevant
features  
of lensing associated with close-orbit planets.

\smallskip 

\noindent{\bf 1.} 
{Every light curve can be used to
either discover close-orbit planets or else to place quantifiable limits
on the presence of planets orbiting the lens in close  
orbits.}  
This is because the  
region in which the deviations occur are low-magnification regions,
and every  
detectable lensing event exhibits 
low magnification, 
whatever 
peak magnification it achieves. 

\smallskip 

\noindent{\bf 2.} 
When the magnification is $A=1+\delta,$ the corresponding value of $\alpha$ is 
\begin{equation}  
\alpha=0.84\, \delta^{\frac{1}{4}}  
\end{equation}  
Thus, {every interval of the low-magnification part of every light curve
can be studied to either discover or place limits on planets at a specific
projected separation 
$\alpha.$}  
The higher the precision of the photometric measurements, the
smaller the values of $\alpha$ we can probe. 
  
\smallskip 

\noindent{\bf 3.}
For each value of $\alpha,$ the value of $q$ determines the 
size of the region over which perturbations of a given magnitude
are detectable.
\begin{equation}  
L(\alpha, q)=2.5\, \xi\, \Bigg[2\, \pi\, 
\Big(\frac{1}{\alpha}-\alpha\Big)\Bigg]\, 10^{[0.5\, Log_{10}(q) - 0.2]}, 
\end{equation}
$L(\alpha, q)$ is expressed in units of the Einstein radius, and 
the value of $\xi$ depends on the photometric sensitivity and frequency
of sampling.   
The value of $L(\alpha, q)$ can be fairly large. For example, for $\alpha=0.25$
and $q=0.001,$ $L(\alpha, q)=0.19.$ This is the radius of the annulus
around $R_\alpha$ within which the perturbations are potentially
detectable. The source must pass through this annulus both on the way
in toward higher magnifications and as it emerges from the higher-magnification
region. 

\smallskip

\noindent{\bf 4.} Orbital motion increases the probability of detection.
In the case considered in point 3 (just above), the total time spent in
this annulus would be $\sim 0.38\, \tau_{E,1}.$ If, e.g., the Einstein
radius crossing time is $30$~days, the source would spend more than $10$~days
crossing the annulus. We would have a very good chance of detecting deviations
for hot Jupiters with orbital periods smaller than $10$~days, because the
perturbed region would rotate into the path of the source one or more
times.    

\smallskip

\noindent{\bf 5.} 
The deviations from the point-lens 
form will have a magnitude that can be easily
computed by using the formula for $L(\alpha, q)$. 
The point-lens magnification will be that for
$u=R_\alpha=\frac{1}{\alpha}-\alpha.$ The upward and downward deviations
 will have
magnitudes approximately corresponding to the point lens magnifications at
$R_\alpha \pm \frac{1}{2}\, L(\alpha, q).$ Thus, for each $\alpha$, 
we can compute the range of magnifications
expected during a deviation for each $q$, and determine how large $q$ would have to be
in order for a planet to produce detectable deviations, given the
quality of the observations. Alternatively one can decide whether 
more sensitive photometric observations
should be taken, in order to be able to detect a planet with a 
particular value of $q$, hence planetary mass.     

\smallskip

\noindent{\bf 6.} For each value of $\alpha,$ the value of $q$ determines the
duration of the deviation. If we assume that orbital motion dominates, then.  
\begin{equation}  
T_{dev}=\frac{L(\alpha, q)}{2\, \pi\, R_\alpha}\, P_{orb} = 
2.5\, \xi\, P_{orb} 10^{[0.5\, Log_{10}(q) - 0.2]}
\end{equation}
In fact, for close orbit planets, orbital motion is likely to dominate
for all but stars with exceptionally large proper motion. Consider a 
solar-mass star at $125$~pc. Equation 4 tells us that, if the lensed source
is in the Bulge, $\theta_E \approx 8$~mas. If $R_\alpha$ is approximately
equal to $3,$ and if the orbital period is $\sim 70$~days, then the orbital
angular speed of the deviation 
is $\sim 0.7^{\prime\prime}$~yr$^{-1}$, larger than the
angular speeds of all but a handful of stars.  
    
\smallskip 

\noindent {\bf 6.} {\sl Nearby Lenses:} 
If the lens star lies within a kpc or so, it is
likely to be detectable. It may be catalogued, perhaps even by the monitoring
programs that search for evidence of lensing. We have found, e.g., that
$\sim 8\%$ of  all lensing event candidates have 2MASS counterparts, many likely
to correspond to the lens (McCandlish \& Di\thinspace Stefano 2011), 
while $>10\%$ of the lenses producing the events
detected by the monitoring programs are predicted to lie within about a kpc
(Di\thinspace Stefano 2008a, 2008b).   Thus, we may know the spectral type of the lens 
and be able to estimate its mass and distance from us. We may even know its 
proper motion.
This information, combined with the Einstein angle crossing time, allows us to
determine the total lens mass and distance. The wide range of other information
potentially derivable from fits to the lensing light curve, may allow us to
also determine the planet's mass and key features of its orbit.
Thus, if we do find a planet, we can learn a great deal about it,
including its gravitational mass, from the lensing observation. In addition, 
because it is nearby, follow up studies to learn more about this planet and to
search for others orbiting the same star may be possible.
On the other hand,
if we do not find evidence of a planet, we can place quantifiable limits on the 
presence of planets with a well-defined range of properties orbiting a star of
known type.  

Given the importance of what we can learn about planets orbiting nearby stars,
it is important to identify those events with counterparts that may be
nearby stars. Thus, in addition to conducting 
automated searches through 
catalogs for possible counterparts to lensing events, we
can employ {\sl Virtual Observatory} (VO)
capabilities to scan existing images of the area within which the
lensing event occurs. By identifying nearby lens stars, 
we can direct resources toward that subset
of events whose study is most likely to be productive through planet
discovery or, alternatively, through providing opportunities to place
meaningful limits on the presence of planets.     

\smallskip

\noindent{\bf 7.} {\sl Distant Lenses:}  
Events associated with close-orbit planets may also be produced 
when $D_L$ is large, particularly when   
$(D_S-D_L)/D_S << 1.$ In such cases, we may not be able to detect
the central star. We therefore may not have any specific information
about its mass or distance from us.  
The value of $\tau_{E,1},$ fit from the light curve,
provides a relation connecting $M, D_L, D_S,$ and
the transverse speed.  
Beyond this, we may have to resort to statistical arguments based on the
distribution of stars in the Galaxy, to provide further constraints.
The deviations may exhibit periodicity, allowing us to estimate the
orbital period, or the fits to the deviations, combined with
other light curve information, may otherwise
allow us to determine approximate values of $\alpha$ and $q$.   

For large $D_L,$ the Einstein angle can be small, comparable
in size to the magnification features associated with deviations.  
This means that finite-source-size effects can be important. Finite
source size can play a negative role by  
softening and diminishing the short-duration deviations associated with the
presence of planets.
Thus, finite-source-size effects may make it more difficult to 
identify the effects of close-orbit planets orbiting distant stars. 
 If, however, the deviations remain
detectable, then the alteration in their shape produced
by finite-source-size may allow us to
derive the value of $\theta_E.$ While this 
will not entirely break the degeneracy,
it does give an extra relation connecting $M$ and $D_L$ (assuming that $D_S$ is known,
at least approximately).   
It is therefore important to include, in the fits to 
deviations in the low-magnification portions of the light curve,
finite-source-size effects.   

\smallskip

\noindent{\bf 8.} {\sl Blending:} When the source star contributes only
a fraction of the baseline light, then the measured magnification
(i.e., the ratio between the light received 
portion of the event and the baseline light) 
during the low-magnification
portion of the event is actually $A_{measured} = 1 + f\, \delta.$  If, therefore,
 we are not aware of the
blending, we will underestimate the value of $\delta,$ hence $\alpha.$
It is therefore important to include the effects of blending in the
light curve fits (Di\thinspace Stefano \& Esin 1995). 
If the event is studied to search for planets as it
occurs, then it is worthwhile observing it in several filters as it occurs,
to determine the amount of baseline light that is lensed, as a function
of wavelength.

\subsection{General Procedure} 

The simple points listed above lead to an important conclusion:
{\sl every light curve can be used to either place limits on the 
presence of possible close-orbit planets or else to discover them.}  
Furthermore, a relatively straightforward procedure can be
employed to achieve these goals. 
 We begin by discussing the case of catalogued events and
then consider what can be learned from ongoing events.

\subsubsection {Catalogued Events} 

Many of the more than 8500 
candidate events 
already discovered\footnote{See, for example,
http://ogle.astrouw.edu.pl/; 
http://www.phys.canterbury.ac.nz/moa/microlensing\_alerts.html.}  
are well-enough sampled at low magnification to provide fertile hunting
grounds for close-orbit planets. 
Not all of the candidate events correspond to lensing events, but 
those with acceptable lens-model fits should be considered
as strong candidates. The fit provides an estimate of 
$\tau_{E,1},$ which relates the total lens  mass to $D_L,$ $D_S,$
and $v.$ The fit also provides a value for the blending
parameter, $f,$ the fraction of the baseline light provided by the
lensed source. Although multiple values of $f$ can be
consistent with the data, the
degeneracy can be lifted if the peak magnification is higher than about $3$
(Wozniak
\& Paczynski 1997). The
degeneracy is also
broken if the event is observed in a variety of wavebands, even
just a few times, or if we have information about other sources of light
along the direction to the event, such as the lens itself.     

To search for close-orbit planets, we  
must search the low-magnification portions of the light curve
for any upward or downward deviations from the point-lens form.
The value of $\delta$ in the region containing the deviation
provides an estimate 
of $\alpha.$ 
If there are several
points per deviation, a model fit can  
provide an estimate of $q.$    
Whatever the numbers of points per
deviation, we 
search for signs of periodicity in the
wings of the light curves. Repeating signatures of close-planet lensing
are not exactly periodic
(Di\thinspace Stefano \& Esin 2011),
but it is possible to introduce a correction to extract
the correct period (Gao et al. 2011).  An interesting feature of the
near-periodicity, is that it is a transient phenomenon, occurring
in the wings of a light curve. That is, if it is due to a
close-orbit planet, it is not a long-term property of the baseline, nor is 
it necessarily exhibited throughout the event.
An orbital period for the portion of the light curve corresponding
to a particular value of $\alpha$ (hence the projected angular
separation), connects the systems mass with  
true separation.  

The combination of these tests provides a great
deal of information about the mass of the lens system, the mass of the
planet, and the size and orientation of the planetary orbit.    
These quantities are all expressed in terms of $D_L,$
$D_S,$ and $v$. The value of $D_S$ is usually known approximately,
because the source is likely to be located in the dense stellar field
being monitored, often the Bulge, but sometimes the Magellanic Clouds 
or M31. Galactic models can be used to construct a probability
distribution for the values of $D_L$ and $v.$ If, however, the lens is 
a catalogued star, then estimates of the values of $D_L, v,$ and $M$
may already be known. Alternatively, observations taken several years
after the event can resolve the separation between
 lens and source, especially
in those cases in which the lens happens to be nearby. This type of study
has already been done for the event MACHO-LMC-5, for which an HST image
taken 6 years after the event was able to provide a photometric parallax
and measure the proper motion,
allowing the gravitational mass of the lens to be determined (Alcock et al.\, 
2001a).  

When there is no sign of deviations or of deviations
that repeat, then it is possible to place limits on the
orbital period and the value of $q$ 
of any planet that might be in close orbit
with the lens star. This is because it is possible to
estimate the 
length of time the magnification is close to
$\delta.$ This tells us the duration $T_{transverse}$ 
of the interval when deviations
caused by a planet with $\alpha=0.84\, \delta^{\frac{1}{4}}$ 
would have been detectable.
For orbital periods shorter than $T_{transverse}$, there would
be a chance to detect deviations caused by the planet at least once. 
Thus, 
by studying the frequency of sampling during this time, 
we can determine the 
duration $T_{dev,min}$ of the shortest deviation to which 
we would have been sensitive. 
This allows us to
compute the smallest value of $q$ to which the 
observations would be sensitive.  
To quantify limits on the presence of planets, we can run a Monte Carlo
simulation in which we model the planetary system, generate
large numbers of light curves,  and compute the fraction 
of all planets within some range of masses, 
orbital separations, orientations, and eccentricities 
would have been discovered, given the
frequency and sensitivity of the observations.

\subsubsection{Newly Discovered Events}

The present discovery rate of candidate lensing events is roughly $1500$  
per year. The sensitivity to low-magnification is good, as witnessed by
the fact that events with estimated peak magnification smaller than $10\%$
are regularly identified. 
Whereas for events that have already finished, we must rely on whatever data
has already been collected, for ongoing events we have opportunities
to collect as much data as would be needed to discover any
 close orbit planets. 
Fortunately, significant improvements in detection efficiency 
can be achieved with relatively modest changes in the observing plan.

The key improvement would be to ensure regular sampling of the baseline.
If, e.g., we want to be able to catch any up-down-up-down deviation
that lasts for at least $12-24$ hours, we could form a team with telescopes
spanning $\sim 4-6$ time zones, with $2-3$ observations per night in each.
The work already carried out by the monitoring teams would constitute
a significant part of the monitoring we  propose, 
so that only modest additional resources would be needed.  
Ideally, each telescope would be able to reliably identify changes in
magnification at the level of a tenth of a percent. To ensure that the
blending parameter can be measured, we might arrange that at, several times
during the underlying event, $2-3$ different filters are employed.  

Note that, although this pattern of monitoring increases the coverage
normally provided by the OGLE and MOA teams, the increase is
relatively modest compared with the kind of intensive, almost
continuous monitoring that takes place over about a night to find
planets with $0.5\, R_E  < \alpha  < 1.5\, R_E.$ We will therefore
refer to the procedure we suggest  as a moderate increase in monitoring.  
Nevertheless, with more than $100$ events occurring at any given time,  it is 
unrealistic to think that this type of program can be carried out for each.
This means that we must select events for special attention.
Criteria that could be useful include the following.

\smallskip

\noindent {\bf 1.} {\sl High peak magnification:} After a handful of points
have been collected as the event rises from the baseline, it is possible to
begin to predict the peak magnification, $A_{peak}.$ Values of $A_{peak}$ 
greater than about 3 make it easier to reliably determine the
blending parameter from the light curve fit. Since this is important
to determining the value of $\delta$ in the low-magnification wings,
it makes sense to devote special attention to events with predicted
high values of the magnification. Moderate monitoring, like that
described above,
 can be started while the light
curve is still on the rise, after the first $5-10$ points above baseline have
been obtained. It is especially important that the modest increase
in monitoring frequency continue during the decline to baseline to 
ensure that, at least on  one side of the light curve, we 
have ideal time coverage. Note that extreme high-magnification
events are already selected for intensive monitoring near peak 
(Griest \& Safizadeh 1998). We suggest that these events receive, in addition,
monitoring that is not so intensive but which supplements what
is normally done at present, to ensure that deviations near
baseline would be detected.  

\smallskip

\noindent {\bf 2.} {\sl Transient periodicity in the wings:}  
A nearly
periodic signal that becomes detectable as the light curve begins to depart 
from baseline may be a signature of close-orbit planets.
To identify such light curves, checks for periodicity could be made
in a sliding window. Windows with a range of sizes should be considered, since the
characteristic size $L(\alpha, q)$ of the region within which perturbations
can be detected is not known {\it a priori}. 

\smallskip

\noindent {\bf 3.} {\sl Counterpart that could be the lens or lensed source:} 
If there is a
counterpart in a catalog, or else if images of the region reveal 
evidence for a possible counterpart, it could be that the counterpart is the 
lens or the lensed source. It is necessary to check that the association between the
position of the event and the possible counterpart is likely to be real, 
which can be accomplished with a Monte Carlo simulation
(McCandlish \& Di\thinspace Stefano 2011). 
Determining whether the counterpart is the lens (making it possible to 
search for nearby planets) or the lensed source (possibly making it
easier to measure the magnification, especially if the baseline is bright) 
can be accomplished through measuring the blending parameter. 

\smallskip
 
\noindent {\bf 4.} {\sl Bright baseline:} Whether or not there is an identified
counterpart, a bright baseline may signal that either
the lensed source or else the lens itself is bright. In the first case  
is ideal for the detection of deviations, and smaller telescopes may be
able to play an important role in monitoring the event. In the latter case, we may have an
opportunity to test for the presence of nearby planets.

\subsection{Close-Orbit Planets, Wide-Orbit Planets and Planets in the ``Resonant Zone''}

Figure 1 demonstrates that planetary systems exhibit a wide
range of separations. From the perspective of gravitational
lensing, they range from ``close'' to ``wide''. The intermediate range,
bounded by the dashed lines, is sometimes called the ``resonant
zone''. It is in this range that planet-lens light curves sometimes
exhibit caustic crossings, and this is the range that on which most
lensing-event searches have concentrated.  
The work we have done in this paper can significantly help with the discovery
of just over $1/3$ of the close-orbit planets, roughly corresponding to
those in orbits with $\alpha > 0.15.$ 

The wide-orbit planets populate 
the upper portion
of Figure~1. Di\thinspace Stefano \& Scalzo (1999b), showed 
that the probability
that an event would ``repeat'', with one portion due to lensing by the star,
and another short-duration portion showing evidence of the planet,  
could be as high as
a few percent to about $10\%.$ 
The type of moderately intensive monitoring we have suggested
for the wings of the light curve to discover close-orbit planets
is also ideally suited to the discovery and study of repeating events. 
The type of monitoring we suggest therefore provides 
opportunities to discover planets in two orbital ranges.
Furthermore, because many wide-orbit planets are likely to be massive enough to  
produce events that last a day or more, failure to detect planets is meaningful.
If, for example, the Einstein-crossing time for a solar-mass planet is 30 days, then 
a $> 2\%$ deviation will typically take $180$~days; a $> 2\%$ deviation
for a Neptune-mass planet would take about $1.3$~days and would be easily 
detected by any program designed to discover close-orbit planets.  

Multiple planets orbiting a single star appear to be common, and one such has 
already been  discovered via lensing (Gaudi et al.\, 2008). In that case, both
planets were located in what we have called the zone for resonant lensing.  
It is certainly possible that we could use lensing to discover planetary systems 
containing both close-orbit and wide-orbit planets. Should this happen, the
signatures of both could appear together in one of the wings of the light curve,
making it important to include both effects when modeling the light curve. 
 
It seems likely that many planetary systems have planets in all three
zones. This suggests an interesting possibility. At present, the search
for planets in the resonant zone focuses on events with extremely high 
values of the magnification. The great advantage of these events is that
intensive, nearly continuous monitoring near peak will
either discover planets or place limits on the existence of planets in
the resonant zone.  We suggest that all such high-magnification events should be
targeted for 
continued monitoring at the moderate intensity required
for close-orbit planets. 
We will thereby either discover close-orbit planets or else place 
quantifiable limits on their existence as well. Furthermore, this 
search has the ability to 
discover wide-orbit planets as well.     
If this plan is followed it will almost certainly lead to the discovery of
interesting systems containing planets in all three zones, and a better understanding of
planetary systems in general.

\subsection{Prospects} 

Searches for close-orbit planets can begin with studies of existing data.
The study of the archived light curves can provide evidence for
planets. Whether or not such evidence is found, these studies can place
limits on the existence of such planets. 
Although the limits might only extend across a limited range of values of $\alpha$
and $q$, they will be new and interesting.

As new data is taken, searches for close-orbit planets can be incorporated without
making major changes. The few changes that will be useful are to 
conduct searches for transient periodicity, and to select some events for
the moderate increases in monitoring near baseline that can improve the chances for
discovering or placing limits on close-orbit planets.    

Ongoing monitoring has already established that events can be identified, and that
binary and planet-events can be found. If there is a new technical challenge
in the study of close-orbit planets, it is posed by the importance
of the low-magnification portion of the light curves. For close orbit planets
we want to measure the value of $\delta= A-1.$ 
Unidentified blending can interfere with these measurements. Yet, it is very
interesting to consider cases in which blending may occur, because nearby stars
which serve as lenses may contribute light to the baseline. Measurements of the blending
parameter are therefore crucial. It is also important to be aware of low-level variability
in light from the stars along the line of sight, including the source and, 
possibly, the lens.
Stellar variability is, however, unlikely to exhibit the form seen in the lower-left panels
of Figures 3, 4, and 5. 

The search for close-orbit and wide-orbit planets can 
proceed simultaneously, using the same observing strategy and data. Fits 
must make sure to model planets in both types of orbit.  
Of particular importance are extreme-magnification events. As 
pointed out above,   
we can use these events to 
search for evidence of close-orbit, resonant-orbit, and wide-orbit planets
occypying the same planetary system.

Interestingly enough, there are other near-term opportunities to search for evidence of
close-orbit planets. One of them is provided by a predicted close passage between the
high-proper-motion dwarf star VB~10
and a background star. This event, slated to occur during the winter of 2011/2012,
 was predicted by Lepine \& Di\thinspace Stefano (2011).
The signatures of any planets that may be orbiting VB~10 are presented in
Di\thinspace Stefano et al.\, (2011). The form of the prediction
is a probability distribution of possible
times and distances of closest approach between VB~10 and
the background star. For those cases in which the distance of closest approach is 
smaller than approximately $60$~mas, close-orbit planets could be detected.
For example, at $60$~mas, a planet with an orbital separation of $0.012$~AU
and an orbital period of $1.7$~days would produce distinctive signatures, similar to
those shown in Figures 3, 4, and 5.         

In addition, it is likely that the {\sl Kepler} space mission will
either discover or place limits on the existence of close-orbit planets.
{\sl Kepler} samples the light from each of its target stars
every 30 minutes.
(Light from a small subset of the targets are 
sampled every minute.)
{\sl Kepler} can 
detect photometric changes of roughly 50 parts per million for a star of
twelfth magnitude. 
This means that {\sl Kepler} can detect low-magnification events
and can probe the low-magnification portion of all events.
We have shown that there is a high probability that 
{\sl Kepler} will observe $\sim$ a dozen low-magnification
events when either a low-mass object passes in front of one of the $150,000$
target stars monitored by the mission, or when a 
target star passes in front of a background source. 
In fact, target stars with
the highest probability of participating in lensing events are presently 
being monitored (PI: Di\thinspace Stefano)

In the future, studies like those needed to discover
 close-orbit planets will be conducted as part of ongoing monitoring
programs.    
Planned programs, such as KNET, will be ideally suited to the discovery
of close-orbit planets. 
KNET is an ambitious monitoring program approved for funding from the
South Korean government. Observing from locations at several positions
in the Southern Hemisphere, KNET will provide continuous coverage of
a large patch of sky, sampling each region with a cadence of 10 minutes.

Gravitational lensing is becoming an effective tool for planet discovery.
The work presented  here shows that some simple modifications
in the way we monitor events have the potential to increase the 
discovery rate by extending the range of our searches. 
The procedures we suggest are ideally suited to the discovery of close-orbit
planets, and will also discover wide-orbit planets. Undoubtedly, there will be 
challenges in implementing these ideas, as there have been with every method
of planet discovery. Within several years however, it should be possible to
regularly discover close-orbit planets both near and far.  

   
\newpage

\begin{figure*}
\begin{center}
Known Exoplanets 
\psfig{file=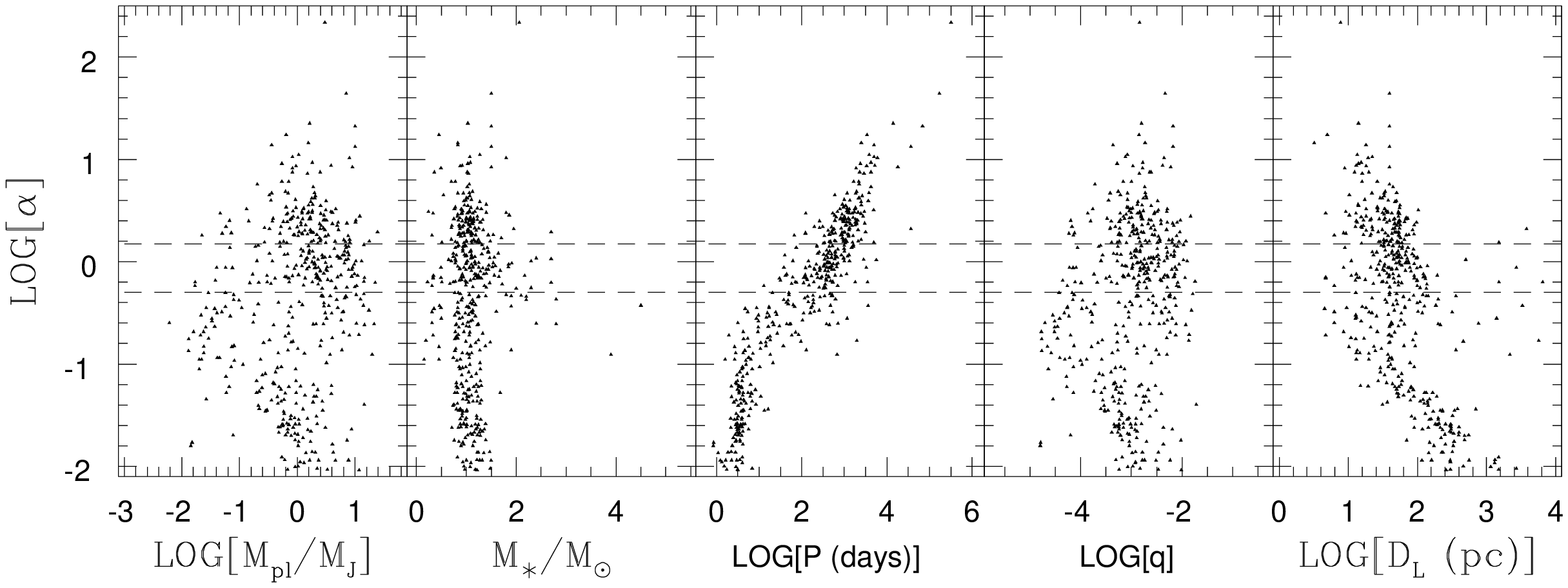,
width=5.0in,angle=-90.0}
\vspace{-2.1 true in}
\caption{
Each point corresponds to a known planetary orbit. In each panel, the value
of $\alpha$ is shown along with a second quantity. From left to right, the
second quantity is the planet's mass, the stellar mass, the orbital period,
the mass ratio $M_{pl}/M_\ast,$ and the distance to the lens.
 Planets in orbits with values of $\alpha$ in between
the two dashed lines are in the ``resonant zone'' and can be found by 
methods already employed. The systems below the lower dashed lines are
the close-orbit
planets we focus on in this paper. Those above the upper dashed line are wide-orbit 
systems. These can produce repeating events that will automatically be discovered
by searches for close-orbit planets.  
}
\end{center}
\end{figure*}

\newpage 

\begin{figure*}
\begin{center}
\psfig{file=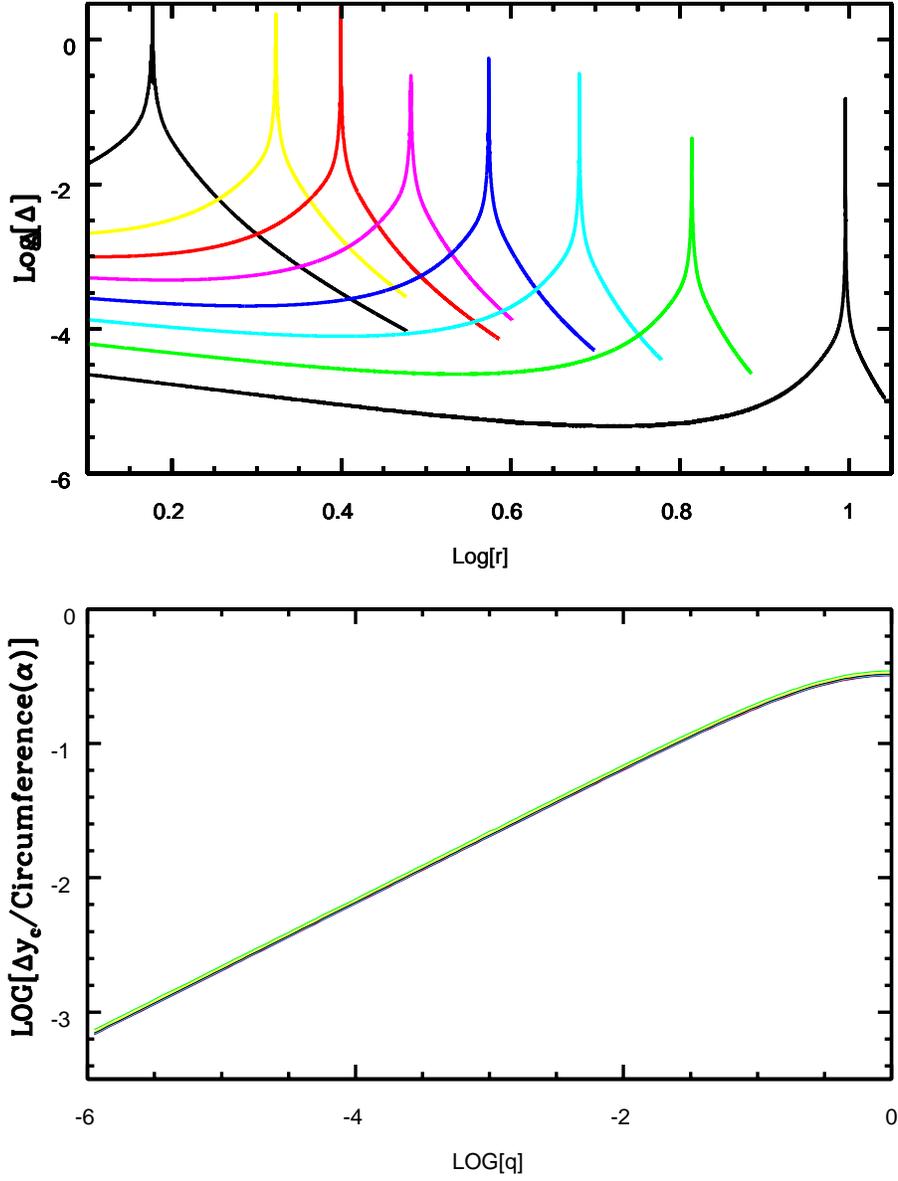,
width=5.0in,angle=0.0}
\caption{
{\bf Top panel:} $log_{10}(\Delta)$ vs $log_{10}(r)$ for a set of planetary systems
with $q=0.001$ and different values of $\alpha.$ 
$\alpha=0.10$ for the right-most curve
and increases by $0.05$ for each curve to the left. $r$ is 
the distance from the center of 
mass; $\Delta$ is the difference between the maximum and minimum magnification
around the ring of radius $r$. For a point lens, $\Delta=0.$     
{\bf Bottom panel:} Normalized distance between small caustic vs $log_{10}(q)$.
The curve is multicolored, with each color corresponding to a different
value of $\alpha$: $0.10$ (blue); $0.15$ (red); $0.20$ (cyan); 
$0.25$ (black); $0.33$ (yellow); $0.40$ (green). The fact that it is difficult to 
resolve these curves shows that there is little dependence on $\alpha.$  
}
\end{center}
\end{figure*}

\newpage 

\begin{figure*}
\begin{center}
\psfig{file=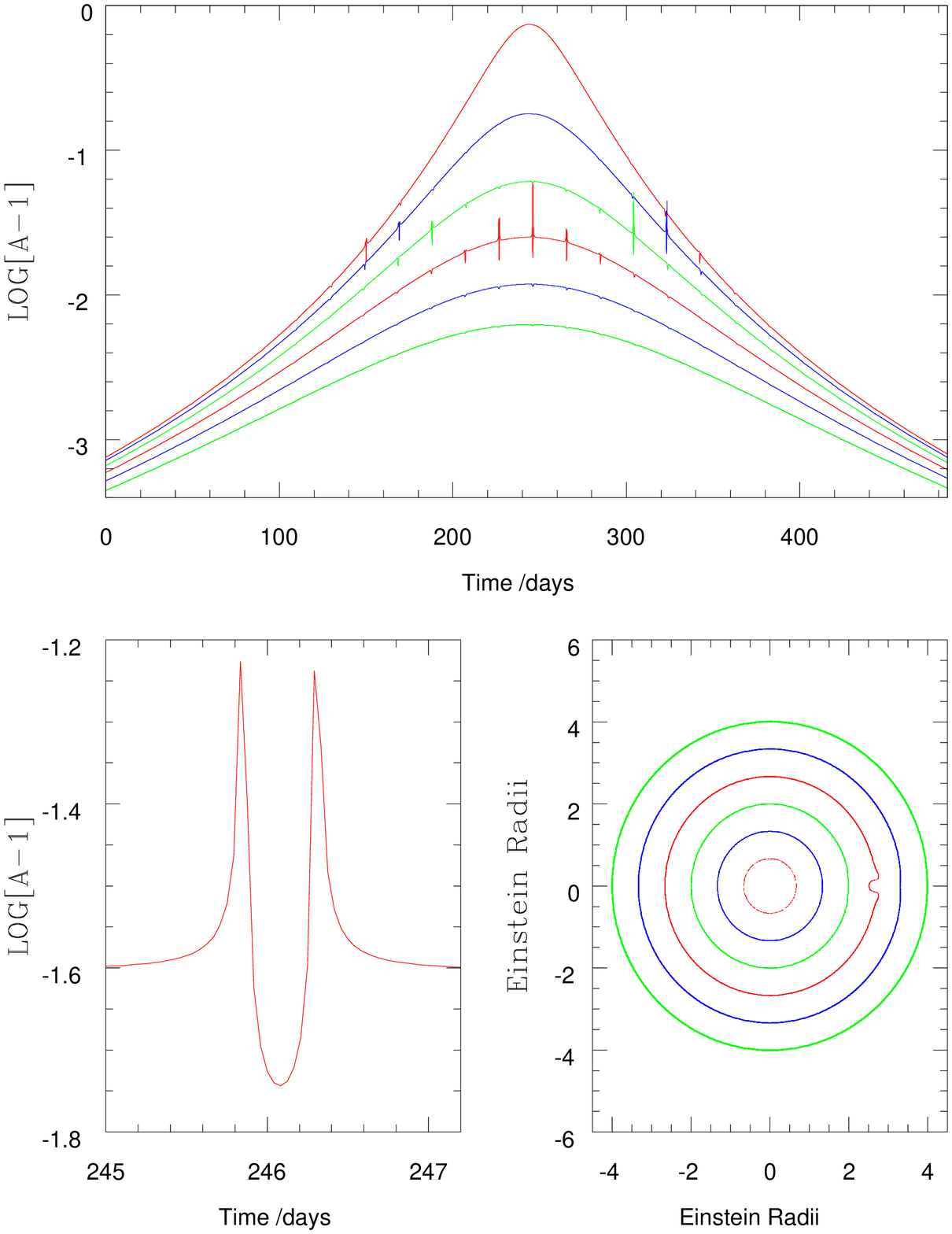,
width=5.0in,angle=0.0}
\vspace{.1 true in}
\caption{{\bf Jupiter-mass planet in orbit with a star of 
$0.8\, M_\odot, \alpha=1/3.$} {\sl Top panel:} light curves. 
Each light curve corresponds to 
a different value of the distance of closest approach: $b=2/3$ in the top curve 
and increases by $2/3$  in each subsequent curve.  
{\sl Bottom left:} Zoomed-in image of a single deviation. 
{\sl Bottom right:} Isomagnification contours associated with the light curves in the
top panel.  
}
\end{center}
\end{figure*}

\newpage 

\begin{figure*}
\begin{center}
\psfig{file=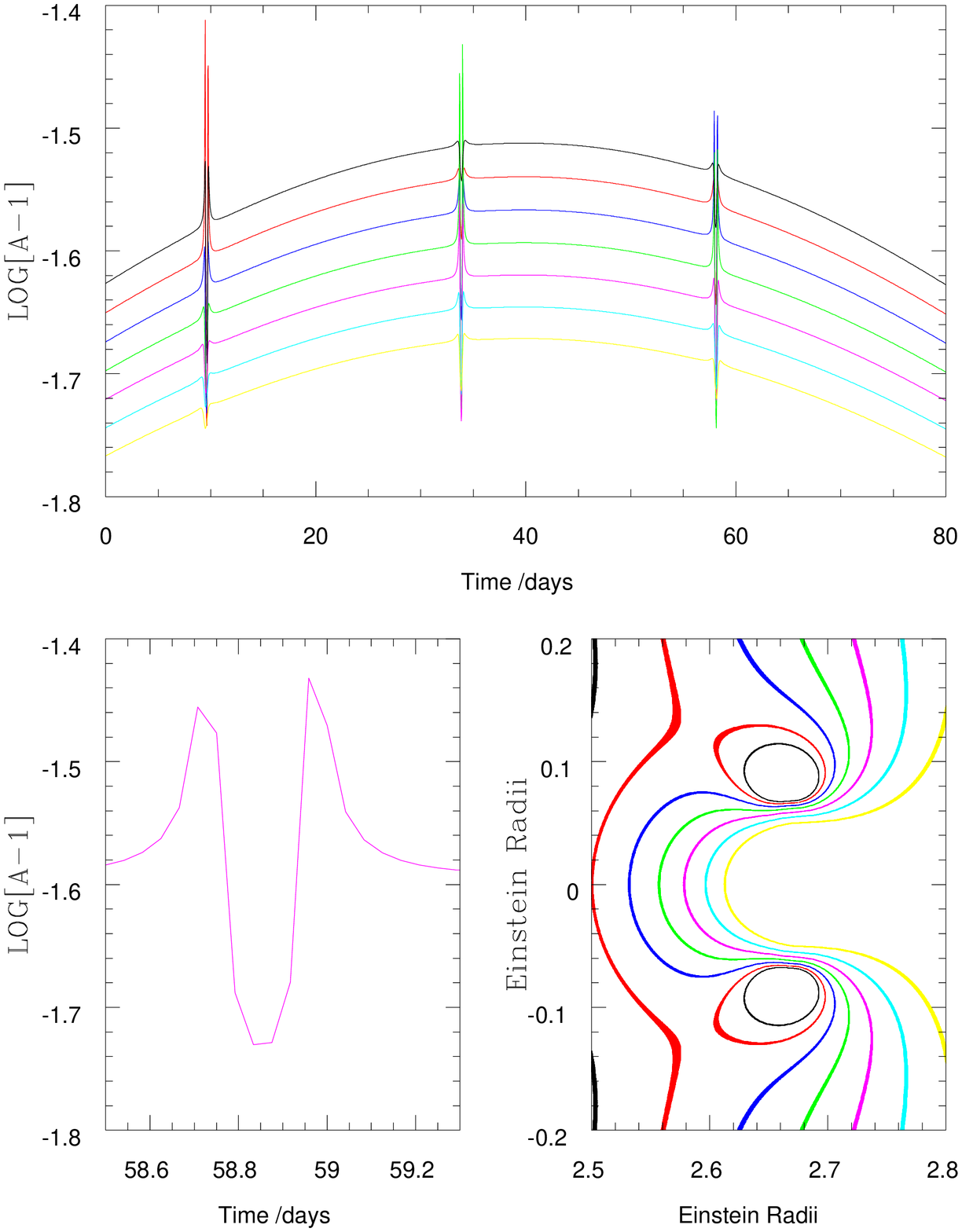,
width=5.0in,angle=0.0}
\vspace{.1 true in}
\caption{{\bf Neptune-mass planet in orbit with a star of 
$0.25\, M_\odot, \alpha=1/3.$} {\sl Top panel:} light curves. 
Each light curve corresponds to 
a different value of the distance of closest approach: $b=2.5$ in the top curve 
and increases by $0.05$  in each subsequent curve.  
{\sl Bottom left:} Zoomed-in image of a single deviation. 
{\sl Bottom right:} Isomagnification contours associated with the light curves in the
top panel.  
}
\end{center}
\end{figure*}

\newpage 

\begin{figure*}
\begin{center}
\psfig{file=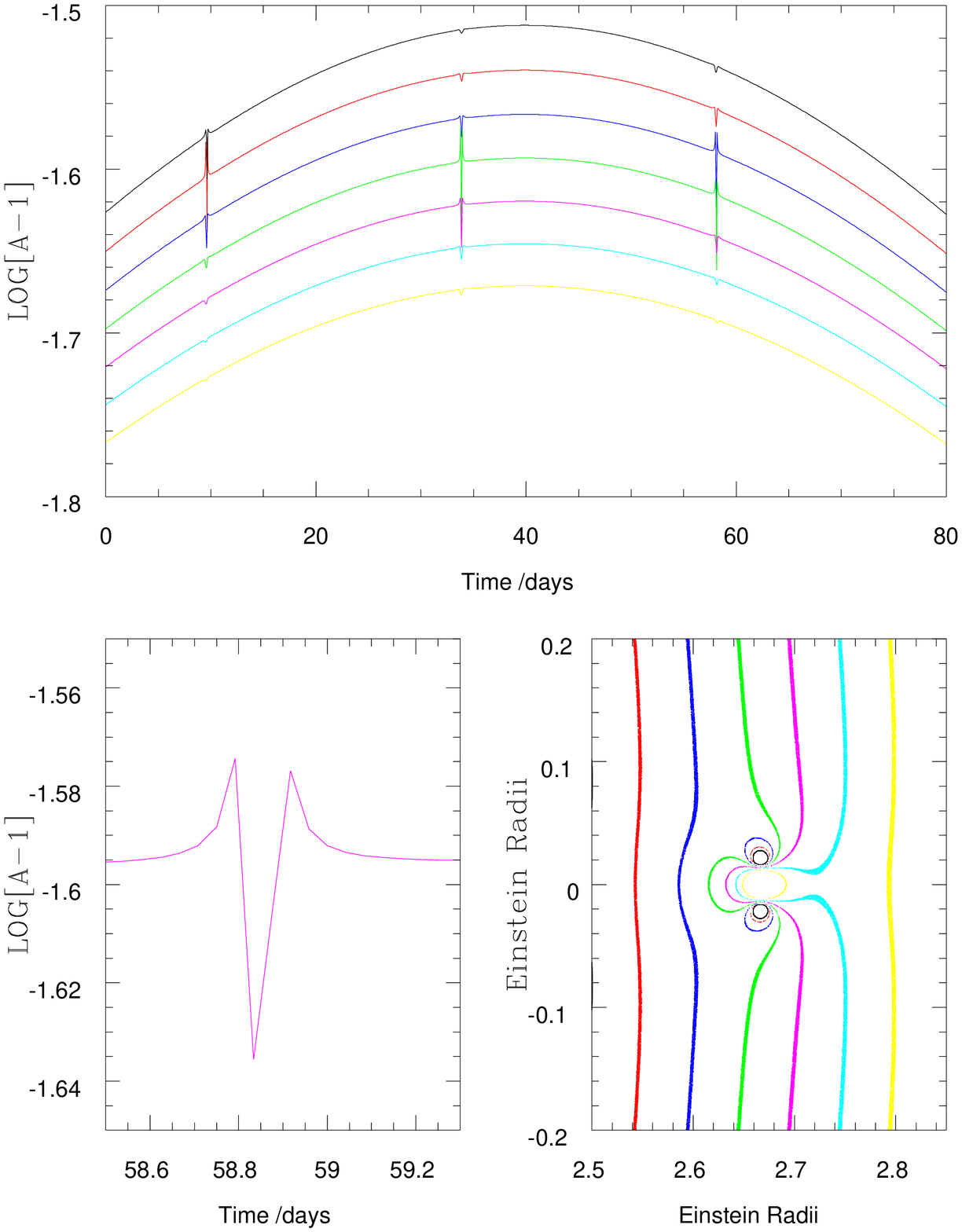,
width=5.0in,angle=0.0}
\vspace{.1 true in}
\caption{{\bf Earth-mass planet in orbit with a star of 
$0.25\, M_\odot, \alpha=1/3.$} {\sl Top panel:} light curves. 
Each light curve corresponds to 
a different value of the distance of closest approach: $b=2.5$ in the top curve 
and increases by $0.05$  in each subsequent curve.  
{\sl Bottom left:} Zoomed-in image of a single deviation. 
{\sl Bottom right:} Isomagnification contours associated with the light curves in the
top panel.  
}
\end{center}
\end{figure*}
\begin{acknowledgements}
I would like to thank Ann Esin and James Matthews for conversations and for 
help with the figures. This work was supported in part by NSF under 
AST-0908878. 
\end{acknowledgements}


\end{document}